\documentclass[twocolumn, tighten]{aastex63}
\usepackage{bm}
\usepackage{amsmath} 
\usepackage{booktabs}
\usepackage{commath}
\usepackage{array}
\usepackage{enumerate}
\usepackage{paralist}

\newcommand{\refereeA}[1]{#1}
\newcommand{\refereeB}[1]{#1}

\newcommand{\KIAA}{\affiliation{Kavli Institute for Astronomy and
Astrophysics, Peking University, Beijing 100871, China}}
\newcommand{\DoA}{\affiliation{Department of Astronomy, School of Physics,
Peking University, Beijing 100871, China}}

\newcommand{\NAOC}{\affiliation{National Astronomical Observatories,
Chinese Academy of Sciences, Beijing 100012, China}}
\newcommand{\BNU}{\affiliation{Department of Astronomy, Beijing Normal 
University, Beijing 100875, China}}

\received{June 1, 2022}
\revised{June 1, 2022}
\accepted{June 1, 2022}
\submitjournal{ApJ}
\shorttitle{Extended Fisher Matrix}
\shortauthors{Z. Wang, et al.}

\begin{document}

\title{Extending the Fisher Information Matrix in Gravitational-wave Data
Analysis}
\correspondingauthor{Lijing Shao}
\email{lshao@pku.edu.cn}
\author[0000-0002-8742-8397]{Ziming Wang}\DoA
\author[0000-0001-7649-6792]{Chang Liu}\DoA\KIAA
\author[0000-0002-9233-3683]{Junjie Zhao}\BNU
\author[0000-0002-1334-8853]{Lijing Shao}\KIAA\NAOC

\begin{abstract}
The Fisher information matrix (FM) plays an important role in forecasts and
inferences in many areas of physics. While giving fast parameter estimation with
the Gaussian likelihood approximation in the parameter space, the FM can only
give the ellipsoidal posterior contours of parameters and lose the higher-order
information beyond Gaussianity. We extend the FM in gravitational-wave (GW) data
analysis using the Derivative Approximation for LIkelihoods (DALI), a method to
expand the likelihood while keeping it positive definite and normalizable at
every order, for more accurate forecasts and inferences. When applied to the two
real GW events, GW150914 and GW170817, DALI can reduce the difference between FM
approximation and the real posterior by 5 times in the best case. The
calculation time of DALI and FM is at the same order of magnitude, while
obtaining the real full posterior will take several orders of magnitude longer.
Besides more accurate approximations, higher-order correction from DALI provides
a fast assessment on the FM analysis and gives suggestions for complex sampling
techniques which are computationally intensive.  We recommend using the DALI
method as an extension to the FM method in GW data analysis to pursue better
accuracy while still keeping the speed.
\end{abstract}
\keywords{methods: data analysis -- methods: numerical -- methods: statistical
-- gravitational waves}

\section{Introduction} 
\label{sec:Intro}

The Fisher information matrix (FM) has been widely used to forecast and estimate
parameters in many fields of physics. In the gravitational-wave (GW) data
analysis, the FM is a useful tool to characterize the parameter-estimation
performance of GW measurements \citep{Finn:1992wt, Vallisneri:2007ev}.  Using
FM, one could estimate the measurement abilities of different GW detectors,
aiming to give guidance on the construction of GW detectors and a glimpse of
scientific outputs \citep{Cutler:1994ys}.  As the next generation GW
detectors---including ground-based and space-borne ones---come into use, it is
critical and inevitable to discuss what and how accurately can they reach. 
Based on FM, many works have investigated in detail the detection ability of
these detectors and given forecasts on the scientific significance \citep[see
e.g.,][]{Cutler:1997ta,Berti:2004bd, Isoyama:2018rjb, Ruan:2020smc, Liu:2020nwz,
Zhao:2021bjw, Shuman:2021ruh, Liu:2021dcr}.

\refereeB{FM is not the only way to help with GW parameter estimation, but it is
the fastest one.} In a widely-used approach, the parameters of interest are
obtained via the Bayes' theorem, and their posterior distributions can be
expressed as the product of a subjectively-selected prior and a model-dependent
likelihood.  One can evaluate the posterior directly using sampling techniques
such as Monte Carlo Markov Chain (MCMC) and Nested Sampling
\citep{2004AIPC..735..395S,Skilling:2006gxv}.  These methods can give the real
full posterior at desired precision in principle, but also take considerable
computing time and resources. This problem is particularly sharp in GW analysis
where one must deal with the complicated waveform templates. For the dozens of
real GW events detected so far \citep{LIGOScientific:2018mvr,
LIGOScientific:2020ibl, LIGOScientific:2021djp}, people are willing to
investigate their parameters as accurate as possible in spite of computational
cost. However, in the exploration of new detectors with rapidly increasing event
numbers, numerous simulations and forecasts are required and they can not all
rely on these delicate sampling techniques given the limited computing
resources. To some extent, forecasting prefers speed to accuracy. Therefore, the
fast FM approximation comes into consideration. It only needs the first order
derivatives of parameters to the waveform at one particular point, the truth
point, which is known in forecasting studies. In mathematics, the FM gives the
Gaussian part of likelihood which can be expressed easily and analytically, and
the covariance matrix of the parameters is just the inverse of FM
\citep{Cutler:1994ys}. With these properties, the FM is used routinely in
numerous studies of GWs.

While benefiting from the advantages, the FM also pays for its approximation.
With the assumption of Gaussian posterior (or Gaussian likelihood in the
parameter space), FM will by definition lead to ellipsoidal confidence-level
contours, whose principal axes represent parameter degeneracy and the areas of
contours reflect the detection ability. \refereeB{However, the Gaussian
likelihood in the parameter space is exact only when the noises are Gaussian and
the model is linear in the parameters. In GW data analysis, we usually assume
that the noise is stationary and Gaussian, while the GW templates are not linear
in the parameters. In this case, the FM comes from the linear signal
approximation of waveform or equivalently the high signal-to-noise ratio (SNR)
approximation \citep{Finn:1992wt,Vallisneri:2007ev}, which is not always
satisfied in reality.}  Besides, one may encounter (quasi-)singular FM in some
cases, which hardly provides useful information. Anyway, only taking out the
Gaussian part is bound to deviate from the real likelihood, and how to check
whether FM is applicable in the forecasting process is worth investigating
\citep{Vallisneri:2007ev,Shuman:2021ruh}. 

There are several ways to deal with this problem without calculating the costly
real likelihood directly. Selecting appropriate parameters to obtain
approximately multivariate Gaussian likelihood in the transformed parameter
space is one good choice \citep{Joachimi:2011iq}. One can also consider the
higher orders of FM and extend it. When adding the higher-order terms, an
approximation closer to the original likelihood can be obtained, and the
accuracy of FM can also be tested by evaluating the higher-order corrections.
\refereeA{As mentioned above, the FM works well in the high SNR limit
\citep{Finn:1992wt}, so one can expand the likelihood in the small parameter
``1/SNR'' \citep[see Appendix A.5 in ][]{Cutler:1994ys,Vallisneri:2007ev}.  An
efficient semianalytical technique has been developed to acquire the exact 
sampling distribution of the maximum-likelihood (ML) estimator across noise
realizations for any SNR \citep{Vallisneri:2011ts}.} Here we will focus on
another path to higher orders, considering directly the Taylor expansion of the
likelihood, which is also the most straightforward way.

When expanding the likelihood at the ML point---or the
truth point in forecasting---the first non-trivial order is exactly the FM
approximation. Therefore, the easiest way to expand FM is to add the
higher-order terms in the Taylor series. However, if truncated at a certain
order, in general this expansion is not guaranteed to be a true mathematically
valid probability distribution, i.e. one that is positive definite and
normalizable. \citet{Sellentin:2014zta} found that a simple rearrangement of the
terms in the Taylor series can guarantee that the expansion remains a true
probability distribution at every order of derivatives of the model. Based on
this, they proposed a new method, named Derivative Approximation for LIkelihoods
(DALI), to approximate the likelihood. At the leading order, the DALI series
gives the FM approximation, while at any higher order the DALI approximation is
positive definite and normalizable, keeping the properties of a probability
density for the approximated likelihood. With DALI, one could obtain the well-defined
non-Gaussian parts of likelihood beyond the FM. As further described in
\citet{Sellentin:2015axa}, the DALI algorithm is independent of the physical
application, meaning that the method can be applied in any fields of physics,
such as cosmology \citep{Sellentin:2014zta} and gravitational lensing
\citep{Sellentin:2015yea}. However, the DALI methods in early works are
approximations of the likelihood in finite-dimensional data spaces with finite
random variables, while in GW analysis in theory we must deal with the
infinite-dimensional data space, namely the GW strain $g(t)$. Therefore, the
original DALI method, reviewd in Appendix~\ref{Appendix A}, needs to make some
changes to fit the requirements in GW astrophysics. 

In this work, we generalize the DALI method to fit the requirements in GW data
analysis and show how it works in reality. \refereeB{Similar to
\citet{Sellentin:2014zta}, this method can show the degeneracy directions and
regions of the parameter space beyond FM. } Meanwhile, it still only needs the
derivatives at one point, keeping itself much faster than those complex sampling
techniques. To show that DALI is a useful and powerful method in forecasts and
inferences, we apply it to two real GW events, GW150914 \citep{Abbott:2016blz}
and GW170817 \citep{LIGOScientific:2017vwq}. We find that compared to FM, the
posterior distribution obtained by DALI approximation is about 1--5 times closer
to the posterior from sophisticated sampling method, measured by the Wasserstein
distance \citep{zbMATH03429908, 2018arXiv180300567P}. Meanwhile, the time costs
of DALI and FM are still at the same order of magnitude. We establish DALI as a
useful extension of FM.

This paper is organized as follows. In Sec.~\ref{sec:Theory} we introduce the
original DALI algorithm in \citet{Sellentin:2014zta} and derive the
corresponding formulae in GW data analysis. In Sec.~\ref{sec:Methods} we explain
how to use the DALI approximation in forecasts and real data analysis.
Sec.~\ref{sec:Examples} gives two examples and compares the DALI approximation
with the FM and sophisticated sampling methods. Sec.~\ref{sec:Summary} presents
a summary and discusses the application prospects of DALI.

\section{DALI algorithm in GW data analysis} 
\label{sec:Theory}

As described in the Introduction, the DALI algorithm is a powerful and fast
method to obtain an approximated likelihood once the model is given
\citep{Sellentin:2014zta}. The detailed expansion formulas of DALI for finite
random variables can be found in Appendix \ref{Appendix A}. To obtain this, we
will start from the classical multi-dimensional Gaussian likelihood in the data
space,
\begin{equation}\label{Gaussian likelihood}
P(\bm{x}|\bm{\Theta}) = \frac{\exp\Big\{{-\frac{1}{2}
\big[\bm{x}-\bm{\mu}(\bm{\Theta})\big]\bm{M}
\big[\bm{x}-\bm{\mu}(\bm{\Theta})\big]}\Big\}
}{(2\pi)^{N_{\rm{d}}/2}\sqrt{|\bm{C}|}} \,,
\end{equation}
and its corresponding log-likelihood ${\cal L} (\bm{\Theta})\equiv - \log
P(\bm{x}|\bm{\Theta})$ for a set of fixed data $\bm{x}$. Here $\bm{x} \equiv
\{x_i \}$ $(i = 1,2,\cdots, N_{\rm d})$ denotes a realization of $N_{\rm d}$
random variables, $\bm{\Theta} \equiv \{\Theta_\alpha \}$ $(\alpha = 1,2,\cdots,
N_{\rm p})$ denotes $N_{\rm p}$ parameters of interest, $\bm{\mu} \equiv
\{\mu_i(\bm{\Theta})\}$ $(i = 1,2,\cdots, N_{\rm d})$ is the model that gives
prediction of $\bm{x}$, $\bm{C}$ is the parameter-independent covariance matrix
in the data space with dimension $N_{\rm d} \times N_{\rm d}$, $\bm{M}\equiv
\bm{C}^{-1}$ is the inverse of the covariance matrix. It is worth to note that
the likelihood (\ref{Gaussian likelihood}) is strictly Gaussian in the parameter
space only when the model $\bm{\mu}$ is linear in its parameters.

With likelihood (\ref{Gaussian likelihood}) and a prior $p(\bm{\Theta})$, one can obtain the
posterior distribution of parameters using the Bayes' theorem,
\begin{equation}
    P(\bm{\Theta}|\bm{x})\propto p(\bm{\Theta})\cdot P(\bm{x}|\bm{\Theta})\,.
\end{equation}
Then, one can evaluate the likelihood at any parameters analytically or use
sampling techniques to get the global distribution of parameters numerically
\citep{Veitch:2014wba}.

However, when the model becomes complicated, evaluating this multi-dimensional
likelihood can be a computationally costly procedure and a good approximation of
the likelihood will help reduce the cost. \citet{Sellentin:2014zta} found a
systematic method, DALI, to obtain a well-defined approximation of likelihood at
any order. The principle of DALI is to rearrange the Taylor expansion of
likelihood in order of derivatives of the model $\bm{\mu}$. The rearrangement
can be complete. When taking ensemble average, denoted as $\left \langle
\,\cdots \right \rangle$, the derivatives of $\cal{L}$ at the ML point can be
written as \citep{Sellentin:2014zta}
\begin{align}\label{Gaussian derivatives}
    {\left \langle \cal{L}_{,\alpha\beta} \right \rangle }\,\,&=
    \bm{\mu}_{,\alpha}\bm{M}\bm{\mu}_{,\beta}\notag \,,\\
    {\left\langle \cal{L}_{,\alpha\beta\gamma} \right \rangle} \,&=
    \bm{\mu}_{,\alpha\beta} \bm{M}\bm{\mu}_{,\gamma} + \bm{\mu}_{,\gamma\alpha}
    \bm{M}\bm{\mu}_{,\beta} + \bm{\mu}_{,\beta\gamma}
    \bm{M}\bm{\mu}_{,\alpha}\notag \,,\\
    {\left \langle \cal{L}_{,\alpha\beta\gamma\delta} \right\rangle} &=
    \bm{\mu}_{,\alpha\beta}\bm{M} \bm{\mu}_{,\gamma\delta} +
    \bm{\mu}_{,\alpha\gamma}\bm{M} \bm{\mu}_{,\beta\delta} +
    \bm{\mu}_{,\alpha\delta}\bm{M}\bm{\mu}_{,\beta\gamma}\notag\\
    &+\bm{\mu}_{,\alpha\beta\gamma}\bm{M}
    \bm{\mu}_{,\delta}+\bm{\mu}_{,\delta\alpha\beta}\bm{M}\bm{\mu}_{,\gamma} +
    \bm{\mu}_{,\gamma\delta\alpha} \bm{M}\bm{\mu}_{,\beta}\notag\\
    &+\bm{\mu}_{,\beta\gamma\delta}\bm{M}\bm{\mu}_{,\alpha} \,,
\end{align}
and so on. Then, rearranging these terms in order of derivatives of the model
$\bm{\mu}$, one could obtain the DALI expansion in Eq.~\eqref{A4}. 

When calculating the likelihood in the GW data analysis, we need to calculate
the probability of GW signal $g(t)$, when the physical parameters $\bm{\Theta}$
and waveform template $h(t;\bm{\Theta})$ are known. In principle, the strain
contains infinite random variables, essentially in a random process. Assuming
that the noise $n(t)$ is stationary and Gaussian with a zero mean, the
likelihood is \citep{Finn:1992wt}
\begin{equation}
    P\big(g(t)|h(t;\bm{\Theta})\big) \propto e^{-\frac{1}{2}(g-h\,,g-h)}\,,
    \label{GW likelihood}
\end{equation}
where $(u\,,v)$ is the symmetric inner product of two data stream $u(t)$ and
$v(t)$,
\begin{align}\label{inner product}
    (u\,,v)&\equiv2\int_{-\infty}^{\infty}
    \frac{\Tilde{u}^*(f)\Tilde{v}(f)}{S_{{n}}(\left |f\right|)}\dif\! f\notag \\
       &= 4 \Re \int_{0}^{\infty}
       \frac{\Tilde{u}^*(f)\Tilde{v}(f)}{S_{{n}}(f)}\dif\! f \,.
\end{align}
Here $S_n(f)$ is the one-sided power spectral density (PSD) of the noise,
$\Tilde{u}(f)$ and $\Tilde{v}(f)$ are the Fourier transforms of $u(t)$ and
$v(t)$, defined by
\begin{equation*}
    \Tilde{u}(f)\equiv{{\cal{F}}}\{u(t)\} =
    \frac{1}{\sqrt{2\pi}}\int_{-\infty}^{\infty}u(t)e^{-2\pi i ft}\dif t\,.
\end{equation*}

Now we use the DALI method to deal with the likelihood in Eq.~\eqref{GW
likelihood}, which requires that the derivatives of the log-likelihood
${\cal{L}} = -\log P\big(g(t)|h(t;\bm{\Theta})\big)$ have similar form to
Eq.~\eqref{Gaussian derivatives}. Indeed they do have the required form,
\begin{align}\label{GW derivatives}
    {\cal{L}_{,\alpha} }\quad\,&= ( h - g \,,h_{,\alpha} )\notag\,,\\
    \,\,\left \langle \cal{L}_{,\alpha\beta} \right\rangle \ \,&=\left \langle
    (h_{,\beta}\,,h_{,\alpha})+( h - g \,,h_{,\alpha\beta})\right\rangle
    \notag\\
    & = (h_{,\beta}\,,h_{,\alpha})+\left \langle  (
    n\,,h_{,\alpha\beta})\right\rangle \notag\\
    &=(h_{,\beta}\,,h_{,\alpha})\notag\,,\\
   \left \langle \cal{L}_{,\alpha\beta\gamma}\right \rangle \,&= \left \langle
   (h_{,\beta\gamma}\,,h_{,\alpha})+(h_{,\beta}\,,h_{,\alpha\gamma})+(h_{,\gamma}\,,h_{,\alpha\beta})\notag\right. \\
    &+ \left.( h - g \,,h_{,\alpha\beta\gamma})\right\rangle \notag\\
    &=(h_{,\beta\gamma}\,,h_{,\alpha})+(h_{,\beta}\,, h_{,\alpha\gamma}) +
    (h_{,\gamma}\,,h_{,\alpha\beta})\notag\,,\\
    \left \langle \cal{L}_{,\alpha\beta\gamma\delta}\right \rangle &=\ \cdots\,.
\end{align}
Through the strict calculation above, we find that the likelihood~\eqref{GW
likelihood} has the same properties as Eq.~\eqref{Gaussian likelihood}. This is
no coincidence. In fact, if one defines the inner product of two random vectors
$\bm{u}$ and $\bm{v}$ in the data space as $(\bm{u}\,,\bm{v})\equiv\bm{uMv} =
u_iM^{ij}v_j$ in Eq.~\eqref{Gaussian likelihood}, then the likelihood can be
rewritten as
\begin{equation}\label{Gaussian inner product}
    P(\bm{x}|\bm{\Theta})\propto
    e^{-\frac{1}{2}\left[\bm{x}-\bm{\mu}(\bm{\Theta})\,,
    \bm{x}-\bm{\mu}(\bm{\Theta})\right]}\,,
\end{equation}
which has the same form as Eq.~\eqref{GW likelihood}. The inner product in
Eq.~\eqref{Gaussian inner product} is a summation, while in Eq.~\eqref{GW
likelihood} the inner product becomes an integral.

We can also make other comparisons of the two likelihoods and the inner
products. For the two components of the inner product, both $\bm{x-\mu}$ and
$g(t)-h(t)$ represent the noise which leads to uncertainty and probability of
the inferred parameters. The covariance matrix $\bm{C}$ describes the
correlation among the random variables, while in the continuous case it is the
auto-correlation function $R(\tau) \equiv\ \left \langle n(t)n(t+\tau) \right
\rangle $  which takes this role. On the other hand, one can see that in
Eq.~\eqref{Gaussian inner product} the inner product is controlled by the
inverse of covariance matrix $\bm{M}=\bm{C}^{-1}$, so in Eq.~\eqref{GW
likelihood} it is expected be controlled by $R(\tau)^{-1} = 1/R(\tau)$. Note
that Eq.~\eqref{GW likelihood} is defined in the frequency domain and
${{\cal{F}}}\{R(\tau)\} = S_n(f)/2$ with $f \geq 0$ \citep{Finn:1992wt}.
Therefore we expect that the inner product will be controlled by $2/S_n(f)$.
This is indeed what Eq.~\eqref{inner product} gives. \refereeA{Although
Eqs.~\eqref{Gaussian likelihood}~and~\eqref{GW likelihood} can be uniformly
described using the inner product, it is still necessary to derive the DALI
algorithm because the gravitational wave signals are continuous, which is
different from the discrete case in \citet{Sellentin:2014zta}.}

\refereeA{In real GW data analysis, we have finite data points due to the
limited timespan and sampling rate of the data, and this is a discrete version
of the continuous signal. We use discrete summation to approximate the integral
in Eq.~\eqref{inner product}, but one must be aware that it is not the summation
in Eq.~\eqref{Gaussian likelihood}. Certainly, one can use the auto-correlation
function to describe the correlations among sampling points in time domain,
obtain the covariance matrix and write a likelihood like Eq.~\eqref{Gaussian
likelihood}. However, this form is complicated due to the non-diagonal
covariance matrix. When the sampling rate tends to infinity, Eq.~\eqref{Gaussian
likelihood} will become the concise form in Eq.~\eqref{inner product} after some
calculation \citep{Finn:1992wt}. }

In all, the DALI algorithm still works in GW data analysis, guaranteed by the
same algebraic structure underpinning Eqs.~\eqref{Gaussian
likelihood}~and~\eqref{GW likelihood}. Therefore, defining the ML parameters
$\hat{\bm{\Theta}}$ and measurement error $\bm{\theta} \equiv \bm{\Theta} -
\hat{\bm{\Theta}}$, the DALI expansion here gives
\begin{eqnarray}\label{DALI-expansion}
    P\!\!&&(\bm{\Theta}) \propto \exp \left[ -\frac{1}{2} ( {h}_{,\alpha}\, , {
    {h}}_{,\beta}) \theta_\alpha \theta_\beta \right.\nonumber\\ 
    &&-\left ( \frac{1}{2}( {h}_{,\alpha\beta}\, , { {h}}_{,\gamma})
    \theta_\alpha \theta_\beta \theta_\gamma + \frac{1}{8} (
    {h}_{,\alpha\beta}\, ,{ {h}}_{,\gamma\delta})\theta_\alpha \theta_\beta
    \theta_\gamma \theta_\delta \right) \nonumber \\
     &&- \left( \frac{1}{6}( {h}_{,\alpha}\, ,{ {h}}_{,\beta\gamma\delta})
     \theta_\alpha \theta_\beta \theta_\gamma \theta_\delta + \frac{1}{12} (
     {h}_{,\alpha\beta\gamma}\, ,{{h}}_{,\delta\tau}) \theta_\alpha \theta_\beta
     \theta_\gamma \theta_\delta \theta_\tau\right.\nonumber\\
  &&+ \left. \left. \frac{1}{72} ( {h}_{,\alpha\beta\gamma}\, ,{
  {h}}_{,\delta\tau\sigma}) \theta_\alpha \theta_\beta \theta_\gamma
  \theta_\delta \theta_\tau \theta_\sigma \right) + {\cal O} (4) \right] \,.
\end{eqnarray}
Note that ${\cal O} (4)$ here means terms that contain $4^{\rm th}$ or higher
orders of derivatives of the model. \refereeA{The highest order term in
Eq.~\eqref{DALI-expansion} is proportional to $( {h}_{,\alpha\beta\gamma}\,
,{{h}}_{,\delta\tau\sigma}) \theta_\alpha \theta_\beta \theta_\gamma
\theta_\delta \theta_\tau \theta_\sigma = (
{h}_{,\alpha\beta\gamma}\theta_\alpha \theta_\beta \theta_\gamma\, ,{
{h}}_{,\delta\tau\sigma} \theta_\delta \theta_\tau \theta_\sigma) $, which is
positive definite and makes the exponential part of Eq.~\eqref{DALI-expansion}
tend to negative infinity for large $\theta$. Therefore, DALI approximation is
normalizable, keeping the properties of a probability density.}

Once the ML parameters, sensitivities of the detectors, and the derivatives of
the waveform are given, one can immediately obtain the approximate likelihood
for the parameters of interest. The logarithm of Eq.~\eqref{DALI-expansion} is
polynomial, making it easy to calculate. Meanwhile, the terms in the two big
pairs of round brackets of Eq.~\eqref{DALI-expansion} are called
``doublet-DALI'' and ``triplet-DALI'' respectively, giving information beyond
the Gaussian ellipsoids \citep{Sellentin:2014zta}. It is worth to note that,
this formula is applicable to any random process with likelihood and inner
product similar to Eqs.~\eqref{GW likelihood}~and~\eqref{inner product}.

\section{Considerations in Applying DALI to GW events} 
\label{sec:Methods}

To obtain a better approximation of the likelihood, we introduce the following
specifics before using the DALI method in GW data analysis.

As the extension of FM, DALI still keeps some properties of FM
\citep{Sellentin:2014zta}, and the most important one is that the DALI
represents the ``average'' detection ability of detectors, which is also why we
take the data average $\left \langle \,\cdots \right \rangle$ in
Eqs.~\eqref{Gaussian derivatives}~and~\eqref{GW derivatives}.\footnote{In this
work, we use FM in the frequentist approach, which is defined as the data
average of ${\cal{L}_{,\alpha\beta} }$. FM can also be defined in the Bayesian
approach and the differences between these two definitions are discussed in
\citet{Sellentin:2014zta}.} \refereeB{When doing forecasts, one usually chooses
a set of typical parameters as the true parameters and expands the likelihood at
the truth point, after an ensemble average of noises. In this case, the results
of DALI are obtained after ``averaging'' over the ensemble, and this cancels out
all effects of particular noise realizations that will happen in a set of real
data \citep[see e.g. ][]{Zanolin:2009mk}.  This is certainly in the sense of
averaging, as we are more interested in the ``average'' detection capability of
the detector because there is no real signal and a specific noise realization.
But for a specific event (corresponding to a specific noise realization), we
need to obtain the ML point from the data of real GW signal at first. Due to the
noise realization, the ML point will deviate slightly from the truth, and the
likelihood contours will also change accordingly. Therefore, when doing
parameter estimation, the DALI series is not expected to be exact with the real
posterior in a strict sense.  DALI here is an approximation with errors in both
noise and truncation of series.} However, as we will show in
Sec.~\ref{sec:Examples}, if one can find a parameter point close enough to the
truth, the DALI can also give an approximation close to the real likelihood. In
principle, when applied to forecasting, DALI is exact in the meaning of average,
and while applied to measuring, DALI is an approximation.  Nonetheless, in both
cases DALI will improve the results given by FM. 

However, not all the useful properties of FM are shared by DALI. The biggest
disadvantage of DALI is that the likelihood shapes of DALI must be obtained
numerically, while in FM, because of the good properties of the Gaussian
function, we can easily evaluate the $n$-$\sigma$ ($n=1,2,3,\cdots$)
confidence-level contours analytically using the inverse of FM when a uniform
prior is chosen \citep{Sellentin:2014zta}.

Based on the rearrangements of Taylor series, it is very important to choose
which parameters to expand because different parameter expansions have different
convergence behaviors. This matter may be not so important in FM where one only
considers the first derivatives of the model, but in DALI we must investigate
the convergence very carefully according to the model. For example, there are
some mathematically equivalent parameter sets controlling the binary masses: (i)
the two-component masses $\{m_1,m_2\}$, (ii) the chirp mass and the mass ratio
$\{{\cal{M}}_c,q\}$, and (iii) the chirp mass and the symmetric mass ratio
$\{{\cal{M}}_c,\eta\}$. But note that the waveform $h$ is roughly the polynomial
of chirp mass and symmetric mass ratio. Therefore, expanding ${\cal{M}}_c$ and
$\eta$ is expected to have a better convergence. Another example is the
luminosity distance $d_L$, which in parameter estimation could be replaced by
(i) the comoving volume $V_c$, (ii) the comoving distance $d_c$, or even (iii)
the reciprocal of $d_L$, namely $d_L^{-1}$. Because the waveform is proportional
to $d_L^{-1}$, so expanding by $d_L$ will face the jumping of the derivative
signs, while the higher order derivatives of the waveform to $d_L^{-1}$ is zero.
Likewise, the convergence of Taylor series of functions in the form of ``$1/x$''
at expanding point $x_0$ is not fine, especially for a small $x_0$.  Therefore,
in practice we choose parameters to obtain DALI series with fast convergence,
just as \citet{Joachimi:2011iq} used the Box-Cox transformations to obtain an
approximately Gaussian likelihood in the transformed parameter space.

\begin{figure*}[htp]
 \gridline{
 \fig{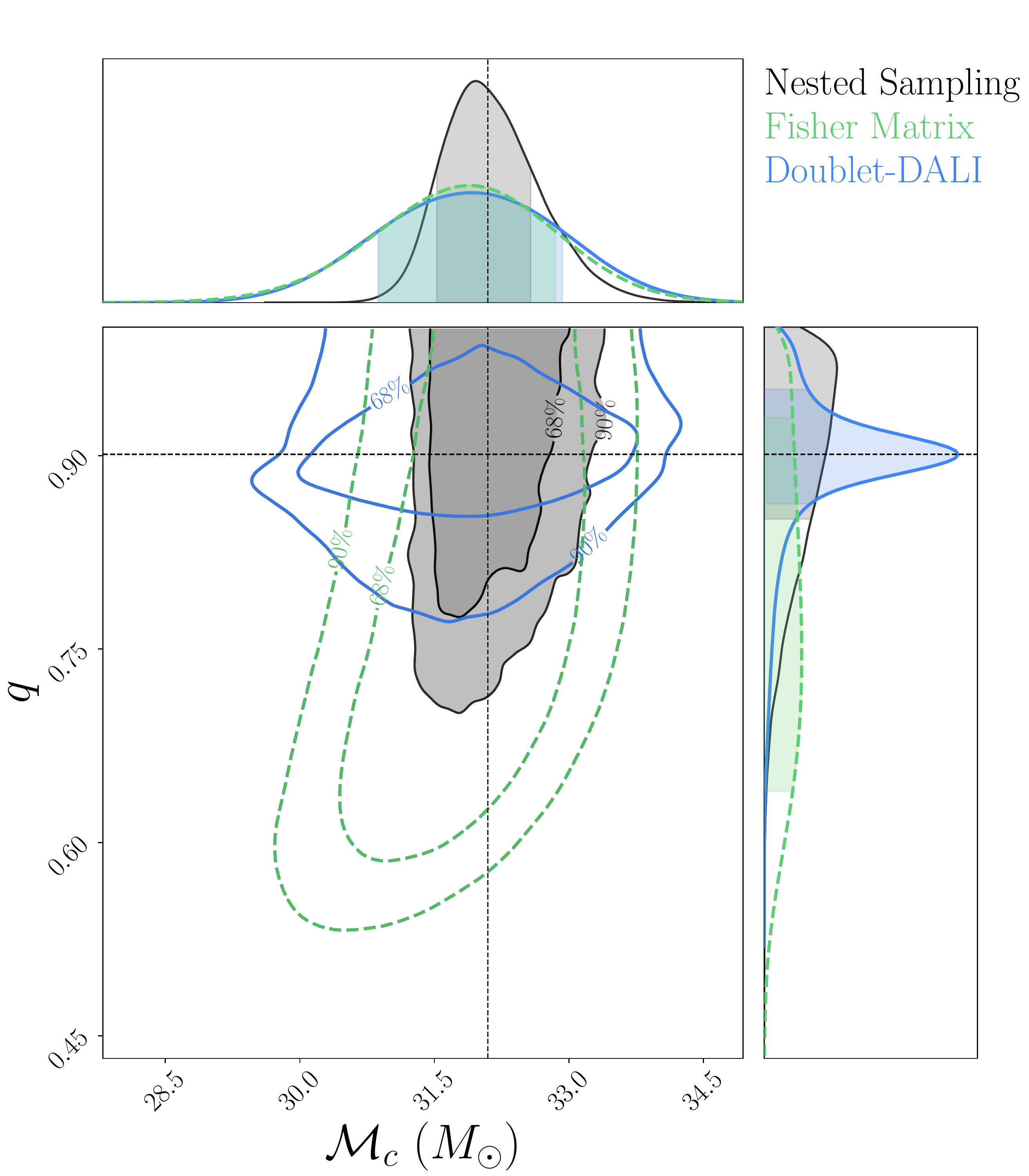}{0.5\textwidth}{}
 \fig{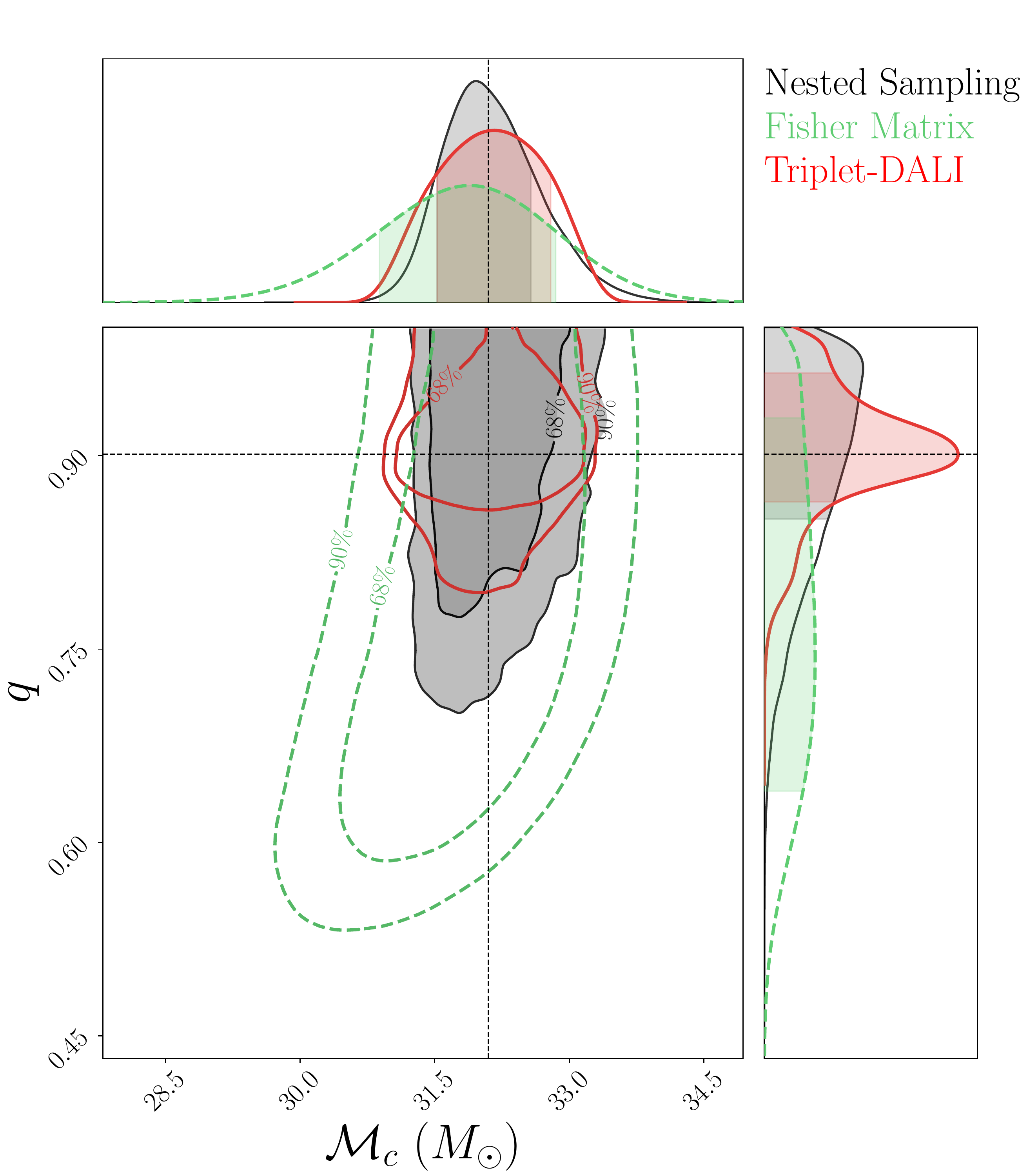}{0.5\textwidth}{}
 }
 \vspace{-0.6cm}
\caption{The $q$-${\cal{M}}_c$ distributions of GW150914 obtained by FM, DALI
approximations and nested sampling. Here, we investigate two kinds of DALI
expansions, doublet-DALI (left) and triplet-DALI (right). In the one-dimensional
marginalized distributions, the medians of nested samplings and 68\% confidence
levels are given by dashed lines and shaded regions, respectively; in contour
plots, 68\% and 90\% confidence levels are illustrated. \refereeA{Note that the
non-elliptical contours given by FM are due to parameter transformation and
truncation of the prior on boundaries.}} 
 \label{fig:150914}
\end{figure*}
\begin{figure*}[htp]
 \gridline{
 \fig{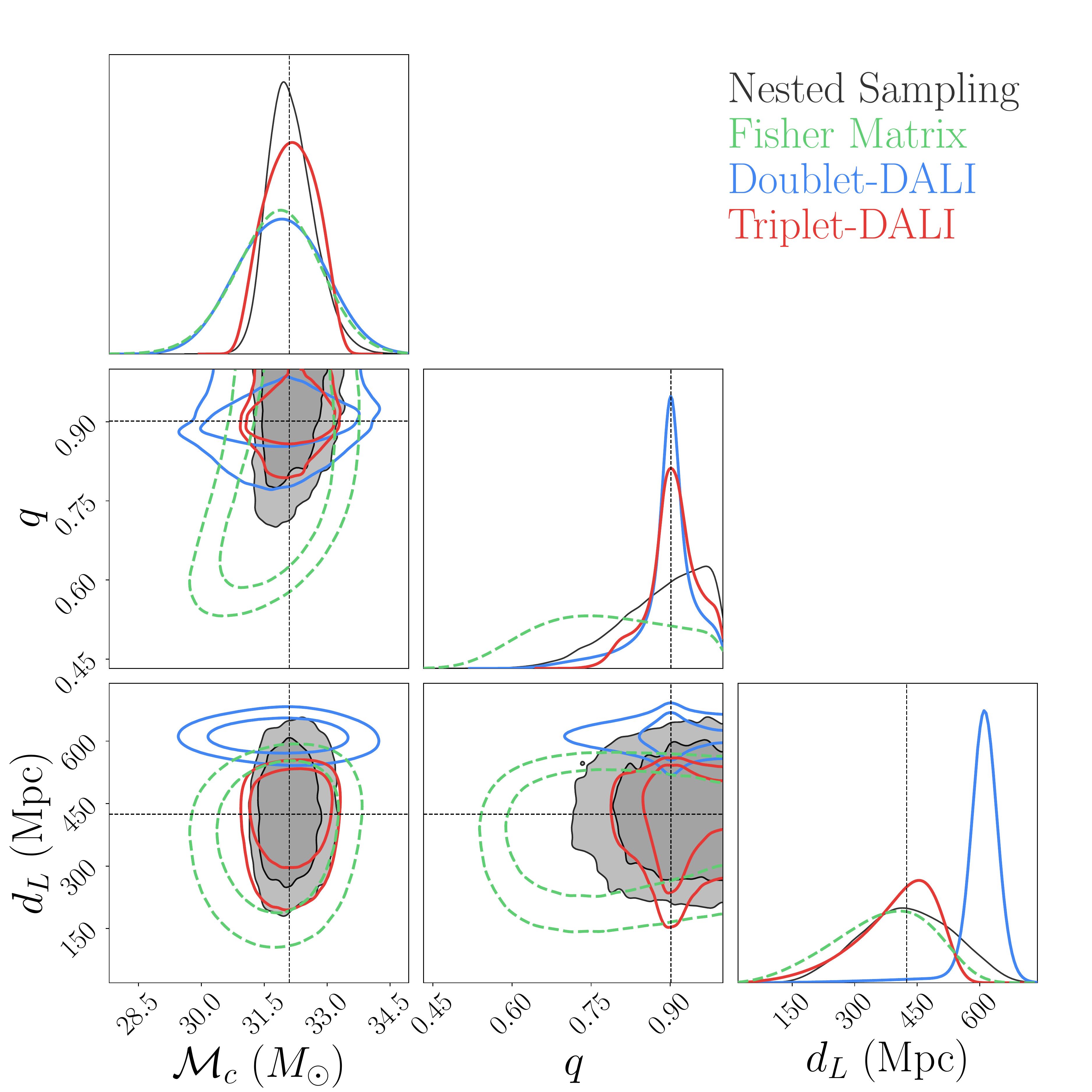}{0.5\textwidth}{\large{GW150914}}
 \fig{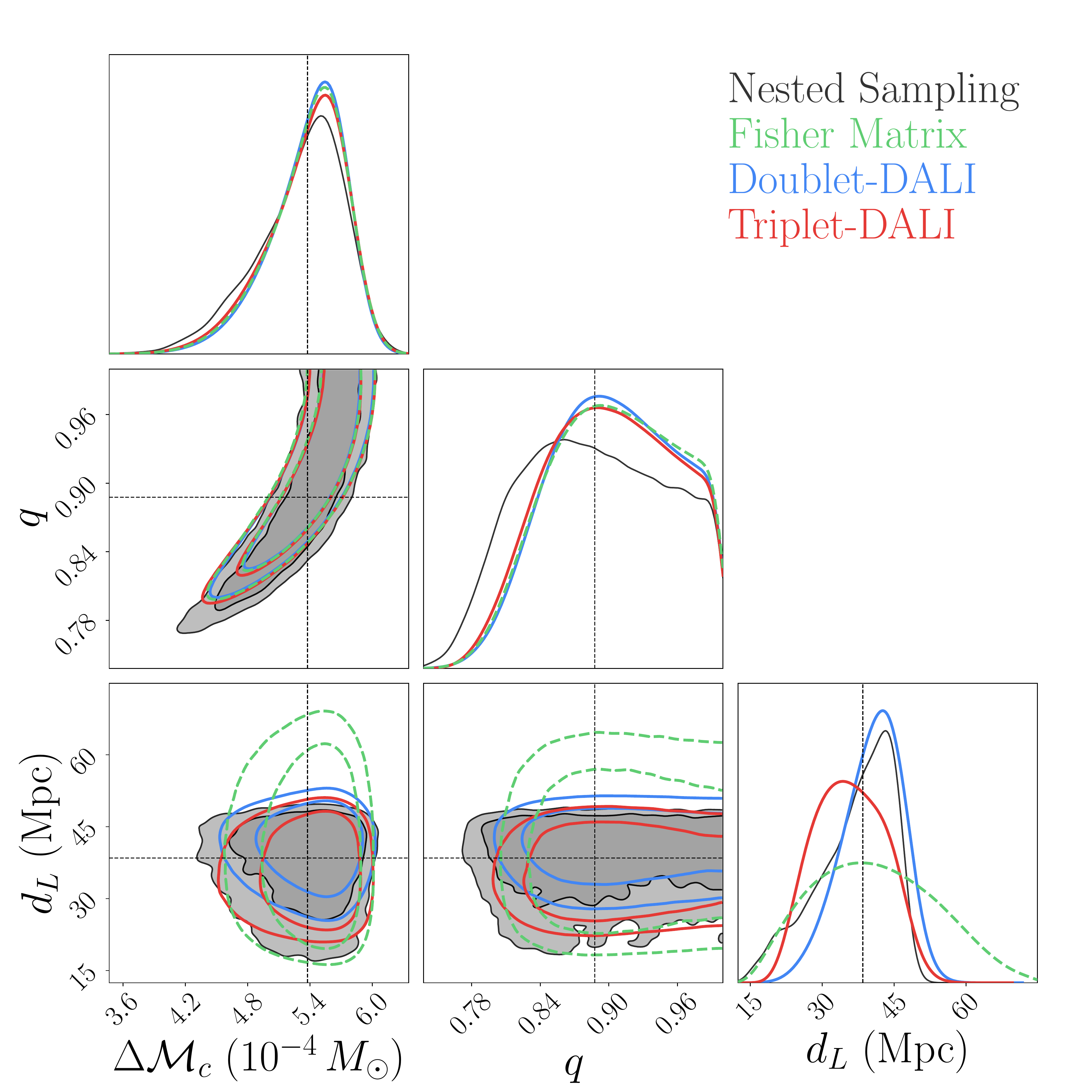}{0.5\textwidth}{\large{GW170817}}
 }
 \vspace{-0.1cm}
 \caption{Same as Fig.~\ref{fig:150914}, but for parameters ${\cal M}_c,q,d_L$
 of  GW150914 (left) and  GW170817 (right). Note that in the right panel, we
 have used $\Delta {\cal M}_c \equiv {\cal M}_c - 1.19700 \, M_\odot$.}
 \label{fig:DALI}
\end{figure*}

The selection of parameters in the likelihood changes the specific functional
form of the prior. Some parameters can provide convenience for derivation, but
lead to complex and extreme priors at the same time. In GW analysis, we usually
choose the prior of the comoving volume $V_c$ to be uniform, meaning that GW
events are uniform in the comoving space. If one chooses the equivalent
parameter $\beta\equiv d_c^{-1}$ (approximately equal to $d_L^{-1}$ when the
redshift $z$ is small) for the convenience of derivation, the prior will not
keep uniform and become  $(\text{d}V_c/\text{d}\beta) \propto \beta^{-4}$. When
$\beta$ is small, this prior dominates the posterior and the shape of likelihood
will be difficult to depict. The influences of prior also exist in FM, and a
detailed discussion can be found in \citet{Vallisneri:2007ev}. 

In all, appropriate selection of expansion parameters can help us reach a better
approximation when truncating the DALI series at a finite order, and both the
uniformity of prior and the convenience of derivation need to be considered
carefully.

\section{Results} 
\label{sec:Examples}

In this section we will show the results of our DALI analysis. We choose two
famous GW events as examples: the binary black hole (BBH) event GW150914
\citep{Abbott:2016blz} and the binary neutron star (BNS) event GW170817
\citep{LIGOScientific:2017vwq}. We use data obtained from the Gravitational Wave
Open Science Center,\footnote{\url{https://www.gw-openscience.org}} a service of
LIGO Laboratory, the LIGO Scientific Collaboration, the Virgo Collaboration, and
KAGRA.

To evaluate the full posterior, we use the open-source software {\sc Parallel
Bilby} \citep{Ashton:2018jfp,Smith:2019ucc,Romero-Shaw:2020owr} and the nested
sampling package {\sc Dynesty}
\citep{2004AIPC..735..395S,Skilling:2006gxv,Speagle:2019ivv} to implement
parameter inference for the real strain data. For the waveform templates, we use
{\sc IMRPhenomD} \citep{Husa:2015iqa,Khan:2015jqa} for GW150914 and {\sc
TaylorF2} \citep{Buonanno:2009zt} for GW170817. The derivatives of these two
waveforms are easy to obtain, and choosing other waveforms will not change our
results significantly.

For the FM and DALI approximation, we expand the likelihood at the medians of
the marginalized distributions from the nested sampling results, because these
median values are expected to be close to the truth values. These examples have
two meanings. On the one hand, one can take the real event as a ``simulation''
in the natural laboratory, and we are forecasting the results via the truth.
This reflects the forecasting ability of DALI. On the other hand, the results
show how good DALI can do as long as we know where to expand the likelihood. In
practice, the expansion point can be obtained independent of the detailed
sampling results, because one only needs a reliable ML point, which is still
much easier and faster to obtain than the full posterior, for example, via a
steepest descent method. 

Both GW150914 and GW170817 have 10 variable parameters. In the parameter
estimation of GW150914, we use chirp mass ${\cal{M}}_c$, mass ratio $q$, two
aligned dimensionless spins $\chi_1$ and $\chi_2$, right ascension $\alpha$ and
declination $\delta$, polarization angle $\psi$, inclination angle $\iota$,
luminosity distance $d_L$, and trigger time $t_c$. For GW170817, we set the
spins of the BNS to be zero, as one would approximate at the leading order from
astrophysics, but add the tidal deformability parameters $\Lambda_1$ and
$\Lambda_2$. The priors of the above parameters are uniform, except $\delta$ and
$\iota$; priors of $\sin \delta$ and $\cos \iota$ are uniform as in common
practice. When expanding the likelihood, we choose the parameter set
$\{{\cal{M}}_c,\eta,\chi_1,\chi_2,\alpha,\delta,\psi,\iota,V_c,\Delta t_c\}$ for
GW150914 and $\{{\cal{M}}_c,\eta,\Lambda_1, \Lambda_2,\alpha, \delta,\psi,
\iota,V_c^{-1/3},\Delta t_c\}$ for GW170817, where $\eta$ is the symmetric mass
ratio, and $\Delta t_c$ is just $t_c$ minus a constant for calculation
convenience. As explained in Sec.~\ref{sec:Methods}, using $V_c$ instead of
$d_c^{-1}$ to expand the likelihood for distant events can prevent the
undesirable features from a non-uniform prior, while for nearby events  the
other choice is better because of the convergence of DALI series. 

After calculating the derivatives at the expansion point, we use the {\sc Emcee}
package \citep{Foreman-Mackey:2012any} to obtain the approximate posteriors of
FM and DALI numerically. The final results are shown in
Figs.~\ref{fig:150914}~and~\ref{fig:DALI}. For the convenience of analysis and
discussion, we only show contours of three important parameters
$\{{\cal{M}}_c,q,d_L\}$. Behaviors of other parameters are similar. 

Figure~\ref{fig:150914} shows the $q$-${\cal{M}}_c$ results of FM, doublet-DALI,
and triplet-DALI, compared with the accurate full posterior of GW150914.
\refereeA{The non-elliptical contours in $q$-${\cal{M}}_c$ distribution given by
FM are due to parameter transformation and truncation of the prior on
boundaries. } Similar to \citet{Sellentin:2014zta}, DALI  gradually improves the
posterior given by FM. One can clearly see that FM gives contours much larger
than the full posterior, which comes from the degeneracy between $m_1$ and
$m_2$. If one had chosen the priors of the two-component masses to be uniform,
the degeneracy becomes even more serious. Containing higher order effects,
doublet-DALI reduces the broad distribution of $q$ given by FM and keeps the
shape of distribution of ${\cal{M}}_c$. The triplet-DALI not only cuts down the
peak of $q$ in doublet-DALI, but also gives a distribution closer to the full
posterior of ${\cal{M}}_c$. 

Figure~\ref{fig:DALI} shows the posterior given by FM, doublet-DALI,
triplet-DALI and nested sampling of GW150914 (left) and GW170817 (right). In
this figure there are many interesting results that reflect the properties of
DALI. The $d_L$ distribution of GW150914 given by doublet-DALI is worse than FM
because in GW150914 we choose $V_c$ instead of $d_c^{-1}$ to expand the
likelihood and if we use $d_c^{-1}$ or $d_L^{-1}$ both FM and DALI will be
controlled by the prior. As explained in Sec.~\ref{sec:Methods}, Taylor
expansion of the ``$1/x$'' type function will suffer from the jumps of the
derivative signs. Indeed, doublet-DALI shifts the maximum of the $d_L$
marginalized distribution to a larger value, and when adding one more order the
triplet-DALI makes the maximum smaller, between FM and doublet-DALI, which is
very similar to the oscillation behavior for the Taylor expansion of $1/x$. When
applied to GW170817 DALI has little improvement on FM in ${\cal{M}}_c$ and $q$,
compared to GW150914. This can be explained by the small ${\cal{M}}_c$ of BNS
and long chirp signal in the data, leading to a high SNR and a quasi-Gaussian
likelihood in the mass parameters ${\cal{M}}_c$ and $\eta$. In this case, the FM approximation can give almost the full
posterior (expanded with appropriate parameters), and DALI  only makes tiny
corrections to it. But for $d_L$, DALI's improvement on FM is obvious,
benefiting from choosing $V_c^{-1/3}$ to expand the likelihood. If one had
chosen $V_c$ instead, a more serious shift like GW150914 in the doublet-DALI
occurs because of the small $d_L$ here. These behaviors are all consistent with
discussions in Sec.~\ref{sec:Methods}. 

To quantify the DALI's improvements on FM, we calculate the Kullback-Leibler
(KL) divergences \citep{kullback1951information} and Wasserstein distances
\citep{zbMATH03429908,2018arXiv180300567P} of the 10 one-dimensional
marginalized distributions between approximates and the full posterior. Both KL
divergence and Wasserstein distance are distance functions defined between two
probability distributions that measure the similarity of them. Compared with the
more commonly used KL divergence, Wasserstein distance is more universal and can
reflect the similarity between two distributions which overlap very little.
Intuitively, if one regards each probability distribution as an equal
(normalized) pile of soil, then the Wasserstein distance can be interpreted as
the required minimum ``work" to deform one pile of soil into another, which is
equal to the amount of soil to be moved multiplied by the average ``distance''
between two piles (distributions). Similar distributions will have a small
Wasserstein distance. 

We show the Wasserstein distances in Table~\ref{table:GW150914:GW170817}.  The
KL divergences between the parameter distributions of GW150914 and GW170817 are
generally consistent with the Wasserstein distances.  In these tables, we can
clearly see DALI's improvements to FM. For example, the triplet-DALI reduces the
difference between FM approximation and the real posterior in the chirp mass
${\cal{M}}_c$ distribution of GW150914 by a factor of 5. As the first order
correction after FM, doublet-DALI is usually radical, giving generally good but
sometimes bad results. Triplet-DALI is more conservative relatively in some
parameters such as $\chi_1$ and $\chi_2$, but it improves FM's results more
comprehensively. Among all parameters, the sky location parameters $\alpha$ and
$\delta$ are worth noting. For GW150914, DALI makes significantly better results
than FM, while for GW170817 DALI is not much different from FM. This can be
explained by the fact that we have data from three detectors for GW170817 but
only two for GW150914.  So the former has almost a Gaussian posterior already
while the latter has more room for improvements.  Generally, the improvements of
DALI for GW150914 are better than GW170817, given the lower SNR of GW150914
whose likelihood is more likely to deviate from Gaussian form.

According to Sec.~\ref{sec:Methods}, the results of DALI must be obtained by
numerical calculations, which leads to longer time cost than FM. However, the
priors of the examples here are not uniform, so discarding priors and only using
FM to calculate contours directly leads to additional errors
\citep{Vallisneri:2007ev}. Therefore, we evaluate the approximate posteriors for
both FM and DALI numerically, which mathematically equates to replacing complex
likelihood using waveforms and noise-weighted inner product with the trivial
polynomial in Eq.~\eqref{DALI-expansion}. In practice, DALI indeed spends more
time than FM, but the calculation times of DALI and FM are still at the same
order of magnitude, while obtaining the full posterior will take several orders
of magnitude more time than FM and DALI. Overall, DALI gives improved
performance, obtaining a posterior closer to the real full posterior while not
costing much more time than FM. In particular, the triplet-DALI algorithm
performs very well in exteding the FM.

\begin{table}[t]
\def\arraystretch{1.2}
\caption{The Wasserstein distances of the 10 one-dimensional marginalized
distributions between FM/DALI and nested sampling for GW150914 and GW170817,
taking the standard deviations of parameters in nested sampling as units.}
\centering
 \vspace{-0.1cm}
\begin{tabular}{cccc }
\hline\hline
Parameter & Fisher &\ \  Doublet-DALI $\ $ & Triplet-DALI \\
\hline
\multicolumn{4}{c}{GW150914} \\
\hline
${\cal{M}}_c$&0.776&0.767&0.157\\
$q$&1.34&0.358&0.356\\
$\chi_1$&8.07&3.79&4.61\\
$\chi_2$&8.23&3.71&4.90\\
$\alpha$&0.614&0.217&0.160\\
$\delta$&1.09&0.677&0.438\\
$\psi$&0.389&0.326&0.367\\
$\iota$&0.22&0.915&0.241\\
$d_L$&0.534&1.64&0.258\\
$\Delta t_c$&0.452&0.587&0.611\\
\hline
\multicolumn{4}{c}{GW170817} \\
\hline
${\cal{M}}_c$&0.165&0.18&0.137\\
$q$&0.267&0.262&0.219\\
$\Lambda_1$&0.254&0.272&0.249\\
$\Lambda_2$&0.172&0.203&0.206\\
$\alpha$&1.97&1.97&2.14\\
$\delta$&0.727&0.651&0.796\\
$\psi$&0.633&0.637&0.626\\
$\iota$&0.525&0.279&0.298\\
$d_L$&0.564&0.463&0.245\\
$\Delta t_c$&0.515&0.534&0.514\\
\hline
\end{tabular}\label{table:GW150914:GW170817}
\end{table}

\section{Conclusion} 
\label{sec:Summary}

The FM is a useful tool to obtain an approximation for complex models and
likelihoods when doing forecasts and inferences. Benefiting from its high speed
and efficiency, FM is widely used in many fields of physics, especially in the
community of  GWs. Lots of works have made use of FM to estimate the detection
ability of GW detectors for specific sources \citep[e.g.,
][]{Finn:1992wt,Cutler:1994ys, Poisson:1995ef, Cutler:1997ta,Berti:2004bd,
Zhao:2017cbb, Wang:2019ryf, Ruan:2020smc,Kang:2021bmp,Shuman:2021ruh}.  Without
doubt, the FM has greatly contributed to the development of GW studies, which
eventually led to the first real GW event GW150914 \citep{Abbott:2016blz} and
the field of GW astrophysics \citep{LIGOScientific:2018mvr,
LIGOScientific:2020ibl, LIGOScientific:2021djp}. At the same time, the
disadvantages of FM are obvious. FM attempts to approximate any likelihood with
Gaussian distribution, which does not apply well in some cases
\citep{Vallisneri:2007ev,Shuman:2021ruh}. Longing for a more accurate
approximation of the likelihood, we have good reasons to extend the FM.

In our work, we extend the FM in GW analysis using the DALI method proposed by
\citet{Sellentin:2014zta}. Note that the original DALI deals with likelihoods
in finite-dimensional data spaces, so in Sec.~\ref{sec:Theory} we first extend
the DALI method to the continuous case to fit the requirements in GW analysis.
The extended result in Eq.~\eqref{DALI-expansion} is universally valid, which
uses inner product to characterize the likelihood whether the data space is
finite or not. Therefore, this formula can be applied to other random processes
with stationary Gaussian noise and other types of likelihoods as long as their
algebraic structures are the same as Eqs.~\eqref{GW
likelihood}~and~\eqref{Gaussian inner product}.

To illustrate the usefulness of DALI, we choose two famous GW events, GW150914
and GW170817, as examples. The results in Sec.~\ref{sec:Examples} can be
regarded as applications in both forecasting and measuring. Compared to FM, we
find that both doublet-DALI and triplet-DALI can bring obvious improvements on
the posteriors for many parameters, and triplet-DALI has overall better
performance than doublet-DALI according to Table~\ref{table:GW150914:GW170817}.
In the two examples, DALI can reduce the difference between FM approximation and
the real posterior by a factor of 5 in the best case, and the calculation time
of DALI and FM is still at the same order of magnitude. Therefore, we suggest
using the triplet-DALI with appropriate parameters to give stable and reliable
corrections to FM. Just as in \citet{Sellentin:2014zta} and
\citet{Sellentin:2015axa}, DALI can reduce the degeneracy that may exist in FM
(see Fig.~\ref{fig:150914}), which gives a fast and one of the simplest validity
check on the FM analysis in forecasts. \refereeA{Recently, it is possible to
estimate the approximated full posterior for real GW events in seconds
\citep{Chua:2019wwt,Dax:2021tsq} or minutes \citep{Cornish:2021lje}. In this
case, DALI can still help the sampling techniques to determine beforehand the
high probability regions to explore. In forecasts, FM and DALI are still
important, because the number of mock events is very large and there is no real
GW signal.} DALI has proved itself to be a successful extension of FM and a
successful application in GW data analysis, keeping a good balance between FM
approximation and the full posterior in terms of fidelity and computational
cost. 

There are still a few possible places where we can improve DALI's performance.
The most direct one is to consider more higher-order terms, which will
definitely lead to a more accurate posterior. However, the number of derivatives
required increases exponentially as the order increases, so is the computational
cost. Another way is to choose more appropriate parameters. The parameter sets
in Sec.~\ref{sec:Examples} can give good results, but there may be better ones.
The third is about the measurement. In this work we use the expansion points
given by nested sampling, while in measurements using DALI we need to find them
independently. A fast and reliable method to find the ML point will
significantly improve the speed and fidelity of DALI.

As an extension of FM, DALI is a useful approximation for likelihoods and makes
better forecasts and inferences beyond FM. It can fill between the simplest but
low-fidelity  FM method and the high-fidelity but costly sampling method. DALI's
ability to break parameter degeneracy in FM is also helpful in some aspects of
GW analysis.

\acknowledgments

We thank Zihe An and Yi-Fan Wang for discussions, \refereeA{and anonymous referees for helpful comments that improved the paper}.  This work was supported by the National
Natural Science Foundation of China (11975027, 12147177, 11991053, 11721303),
the National SKA Program of China (2020SKA0120300), the Young Elite Scientists
Sponsorship Program by the China Association for Science and Technology
(2018QNRC001), the Max Planck Partner Group Program funded by the Max Planck
Society, and the High-Performance Computing Platform of Peking University.  \refereeA{Z.W.
was supported by the Hui-Chun Chin and Tsung-Dao Lee Chinese Undergraduate 
Research Endowment (Chun-Tsung Endowment) at Peking University}, and J.Z. was 
supported by the ``LiYun'' postdoctoral fellowship of Beijing Normal University.

 \vspace{5mm}
 \facilities{LIGO/Virgo/KAGRA}

\software{Bilby \citep{Ashton:2018jfp},
          ChainConsumer \citep{Hinton2016},
          dynesty \citep{2004AIPC..735..395S,Skilling:2006gxv,Speagle:2019ivv},
          emcee \citep{Foreman-Mackey:2012any},
          Parallel Bilby \citep{Smith:2019ucc},
          PyCBC \citep{alex_nitz_2022_5825666}
          }


\appendix

\section{DALI Algorithm}
\label{Appendix A}

The Derivative Approximation for LIkelihoods (DALI) algorithm was developed by
\citet{Sellentin:2014zta}. A brief discussion on the principle and
implementation of the DALI algorithm is given below. It also sets notations for
the paper. 

The ultimate goal for parameter estimation is to obtain a probability density
function, $P(\bm{\Theta})$, where $\bm{\Theta} \equiv \{\Theta_\alpha \}$
$(\alpha = 1,2,\cdots, N_{\rm p})$ collectively denotes $N_{\rm p}$ parameters
of interest. We use Greek indices to run over the parameters, and Latin indices
to run over the data. By performing Taylor expansion to the log-likelihood,
${\cal L} (\bm{\Theta})\equiv - \log P(\bm{\Theta})$, around best fitting
parameters, $\hat{\bm{\Theta}}$, one obtains \citep{Vallisneri:2007ev,
Sellentin:2014zta}
\begin{equation}\label{eq:Taylor}
    P(\bm{\Theta}) \propto \exp \left[ - \frac{1}{2!}{F^{\alpha\beta}}
    \theta_\alpha \theta_\beta - \frac{1}{3!} S^{\alpha\beta\gamma}
    \theta_\alpha \theta_\beta \theta_\gamma - \frac{1}{4!}
    Q^{\alpha\beta\gamma\delta} \theta_\alpha \theta_\beta \theta_\gamma
    \theta_\delta + {\cal O} \left( \theta^5 \right) \right] \,,
\end{equation}
where $\bm{\theta} \equiv \bm{\Theta} - \hat{\bm{\Theta}}$, and we have defined
\begin{equation}
    F^{\alpha\beta} \equiv \left. \frac{\partial^2 {\cal L}}{\partial
    \Theta_\alpha \partial \Theta_\beta} \right|_{ \bm{\Theta} =
    \hat{\bm{\Theta}} } \,, \quad
    S^{\alpha\beta\gamma} \equiv \left. \frac{\partial^3 {\cal L}}{\partial
    \Theta_\alpha \partial \Theta_\beta \partial \Theta_\gamma}\right|_{
    \bm{\Theta} = \hat{\bm{\Theta}} } \,, \quad
    Q^{\alpha\beta\gamma\delta} \equiv \left. \frac{\partial^4 {\cal
    L}}{\partial \Theta_\alpha \partial \Theta_\beta \partial \Theta_\gamma
    \partial \Theta_\delta}\right|_{ \bm{\Theta} = \hat{\bm{\Theta}} } \,.
    \label{A2}
\end{equation}

We will use shorthand notations for derivatives in the parameter space,
\begin{equation}
    F^{\alpha\beta} = {\cal L}_{,\alpha\beta}\,, \quad
    S^{\alpha\beta\gamma} = {\cal L}_{,\alpha\beta\gamma} \,, \quad
    Q^{\alpha\beta\gamma\delta} = {\cal L}_{,\alpha\beta\gamma\delta} \,,
    \label{A3}
\end{equation}
where they are understood to be evaluated at $\bm{\Theta} = \hat{\bm{\Theta}}$.
The expansion (\ref{eq:Taylor}) is only well-defined as a normalizable
probability density at its second order, characterized by $F^{\alpha\beta}$, and
$F^{\alpha\beta}$ is the usual FM \citep[see e.g.][]{Finn:1992wt,
Cutler:1994ys}. Terms characterized by $S^{\alpha\beta\gamma}$ and
$Q^{\alpha\beta\gamma\delta}$ are not positively defined in general
\citep{Sellentin:2014zta}.

To cure the problem, instead of performing the Taylor expansion, the DALI
algorithm expands $P(\bm{\Theta})$ in the order of derivatives
\citep{Sellentin:2014zta}. Assume a theoretical model $\bm{ {\mu}}(\bm{\Theta})$
whose component ${\mu}_i(\bm{\Theta})$ $(i=1,2,\cdots, N_{\rm d})$ predicts the
$i^{\rm th}$ data point, and a covariance matrix ${\bm{C}}$ with dimension
$N_{\rm d} \times N_{\rm d}$ in the data space whose inverse is $\bm{M} \equiv
{\bm{ C}}^{-1}$. The DALI expansion gives,
\begin{eqnarray}\label{A4}
    P(\bm{\Theta}) &\propto& 
    \exp \left[ -\frac{1}{2} \bm{ {\mu}}_{,\alpha} \bm{M}
    \bm{ {\mu}}_{,\beta} \theta_\alpha \theta_\beta - \left( \frac{1}{2}
    \bm{ {\mu}}_{,\alpha\beta} \bm{M} \bm{ {\mu
    }}_{,\gamma} \theta_\alpha \theta_\beta \theta_\gamma + \frac{1}{8}
    \bm{ {\mu}}_{,\alpha\beta}\bm{M} \bm{ {\mu}}_{,\gamma\delta} \theta_\alpha \theta_\beta \theta_\gamma
    \theta_\delta \right) \right. \\
    && \left. - \left( \frac{1}{6} \bm{ {\mu}}_{,\alpha}
    \bm{M}\bm{ {\mu}}_{,\beta\gamma\delta} \theta_\alpha
    \theta_\beta \theta_\gamma \theta_\delta + \frac{1}{12} \bm{ {\mu}}_{,\alpha\beta\gamma} \bm{M} \bm{ {\mu}}_{,\delta
    \tau} \theta_\alpha \theta_\beta \theta_\gamma \theta_\delta
    \theta_\tau + \frac{1}{72} \bm{ {\mu}}_{,\alpha\beta\gamma}
    M \bm{ {\mu}}_{,\delta\tau\sigma} \theta_\alpha
    \theta_\beta \theta_\gamma \theta_\delta \theta_\tau \theta_\sigma
    \right) + {\cal O} (4) \right] \,, \nonumber
\end{eqnarray}
where the first term in the exponent gives the usual FM, while the following two
pairs of round brackets give the corrections at higher orders, and they are
named respectively as the ``doublet-DALI'' and the ``triplet-DALI''
\citep{Sellentin:2014zta}. \refereeA{Such an expansion guarantees the properties
of probability densities like Eq.~\eqref{DALI-expansion}. }

\bibliography{refs}{}
\bibliographystyle{aasjournal}

\end{document}